\def\etal{{\it et al.~\/}}

\def\ie{{\it i.e.~\/}}

\def\ltsima{$\; \buildrel < \over \sim \;$}
\def\simlt{\lower.5ex\hbox{\ltsima}}
\def\gtsima{$\; \buildrel > \over \sim \;$}
\def\simgt{\lower.5ex\hbox{\gtsima}}

\documentstyle[11pt,paspconf]{article}

\begin{document}

\title{The First Structures in the Universe: Pop III Objects}
\author{Andrea Ferrara}
\affil{Osservatorio Astrofisico di Arcetri, Firenze, Italy}
\author{Simone Marri}
\affil{Dip. di Astronomia, Universit\'a di Firenze}

\begin{abstract}
We review some of the possible observable effects (reionization,
feedback on galaxy formation, supernovae and metal enrichment)
of the first collapsed, luminous (Pop III) objects in the universe. 
We show that supernovae in
these objects should be considerably magnified by the intervening
cosmological matter distribution; the implications of this process
are briefly discussed, anticipating some of the results of Marri \&
Ferrara (1998).

\end{abstract}


\section{Introduction}

As the temperature of the cosmic bath decreases, atoms start to recombine
and therefore decouple from  CMB radiation at redshift $\approx 1100$.
The baryonic Jeans mass after this event is given by (assuming $\Omega =1$) 
\begin{equation}
\label{mj}
M_j \simeq 6\times 10^4 \left({1+z\over 30}\right)^{-3/2} \left({T\over 500 {\rm K}}\right)^{3/2}
\Omega_b M_\odot,
\end{equation}
where $T$ is the gas temperature and $\Omega_b$ the baryon density parameter.
Masses larger than $M_j$ are gravitationally unstable and should, in principle,
collapse. However, in order for the actual collapse to occur a more severe condition 
must be satisfied, \ie that the cooling time of the gas is shorter than the Hubble time
at that epoch; in fact radiative losses provide the only way for the gas to lose
pressure and to settle down in the potential well of the dark matter halo.
Since the virial temperature corresponding to Pop~III object masses is typically
$\simlt 8000 K$, cooling by hydrogen Ly$\alpha$ excitation is strongly
quenched,  and the only viable coolant in a primordial H-He plasma is molecular hydrogen. 
$H_2$ is produced during the recombination phase, but its relic
abundance is so small that it contibutes only marginally to the cooling. However,
its fractional abundance increases during the collapse phase along with the density,
finally becoming dominant. Tegmark \etal (1997) have determined the minimum 
mass necessary for collapse and they find that these first 
objects in a standard CDM model should form at redshift $\approx 30$ and have 
total masses $M\approx 10^6 M_\odot$. 
During the collapse, if fragmentation occurs, a stellar cluster is likely to be formed
with a stellar mass, $M_*$, which depends on the (unknown) details of the star formation
process. An educated guess, based both on the experimental data available
for the Galaxy and on theoretical arguments, suggests that $M_* \approx 10^{2-3}$.
The number of massive stars and supernovae produced depends on the postulated IMF;
if the latter is not too steep, it is reasonable to expect that the Pop III star cluster
will be a source both of ionizing photons and (explosive) energy injection in the surrounding
environment. In the
next section we provide some estimates of these effects and finally we discuss
how these (very) high redshift supernovae might be close to enter the realm of
visibility in the next few years, thanks to their relatively strong magnification due
to the intervening cosmological matter distribution.

\section{Effects of Pop~III Objects}
\subsection{Reionization and Feedback}
As Pop III objects start to shine ionizing photons produced by their pristine
stellar clusters, they will create an approximately spherical ionized (HII) 
region in the surrounding IGM.
The radius of each sphere is relatively small ($\approx 0.1-0.2$~Mpc); this value depends
linearly on the (unknown) value the fraction of ionizing photons
escaping from the Pop III ISM, here assumed to be $f_{esc}\sim 0.2$. Therefore,
the reionization epoch, defined as the $z$ at which the ionized spheres overlap,
is rather delayed with respect to the formation of the Pop III, \ie $z\approx 10$
(Ciardi \& Ferrara 1997).
The above calculation requires that the galaxy formation follows closely the underlying
hierarchical clustering that aggregates dark matter halos. This, however, needs
not to be the case. In fact, in addition to ionizing the intergalactic hydrogen,
the emitted radiation in the soft-UV band $11.2-13.6$~eV is also able to photodissociate
$H_2$ molecules inside collapsing objects, thus temporary inhibiting the formation 
of objects smaller than the mass threshold for the ignition of Ly$\alpha$ cooling
($M \approx 10^8 M_\odot$). This process is often referred to as a kind of
"negative feedback'', according to Haiman \etal 1997, who first suggested that
such mechanism can be at work. More recently, this negative role of Pop III
on subsequent galaxy formation has been revisited by Ciardi, Abel \& Ferrara (1998).
Without entering the details of their argument, the main point is that it appears 
that Haiman \etal (1997) have considerably overestimated the photodissociating
soft-UV radiation produced by Pop III objects, calculated by the crucial assumption
that HII spheres (and consequently the photodissociation spheres) overlap already
at high $z$. Instead, as discussed above,  overlap occurs only 
much later, at an epoch at which the role of $H_2$ cooling in the collapsing galaxies
has become negligible  with respect to atomic cooling, thus  strongly reducing, if not
even suppressing, the alleged negative feedback. 

\subsection{Explosions and Metal Enrichment}

In the likely situation in which stars more massive than $8 M_\odot$ are born in 
the Pop III stellar cluster, supernovae will start to explode after a few million years.
Even if the stars in the cluster are all coeval (\ie born from the same initial burst
of star formation) supernovae will explode at different times due to their mass
spread. For a stellar cluster inside a Pop III with properties as the ones discussed
above, we expect, for a Miller-Scalo IMF, about $N=2-20$ supernovae to blow off.
The total mechanical luminosity of such a multi-SN explosion, approximated 
as a continuous energy injection,  is of the order of  $\epsilon_0 N/t_{_{OB}} 
\simeq .6-6\times 10^{37}$~erg~s$^{-1}$, where $\epsilon_0 \approx 10^{51}$
is the energy of a single supernova explosion and $t_{_{OB}} \approx 10^7$~yr
is the typical lifetime of a massive star association. The point to be appreciated is that
Pop III objects, due to their low mass and binding energy, are very fragile 
objects and tend to be {\it blown-away} (Ciardi \& Ferrara 1997) by explosions:
a number $N_c \simeq M_6^{5/3} (1+z/30) h^{2/3}$ of supernovae will suffice to
disrupt the object, where $M_6=M/10^6 M_\odot$. Since for our reference 
stellar cluster $N > N_c$, its residual gas content
will be swept away by the expanding multi-SN driven shock and lost to
the IGM. As a result, further star formation will be inhibited and the remaining 
low mass stellar population will continue to evolve passively. The final fate 
of this object is not yet clear: stars can become unbound due to dynamical 
instabilities following the blow-away or they can steadily evaporate from the 
stellar cluster and be lost in the field; the timescale of these processes, though, 
should be compared with the one for merging. The gas injected in the IGM
is enriched by the heavy elements produced by supernova nucleosynthesis
processes. Assuming that the mass in heavy elements averaged on a 
Miller-Scalo IMF is $\approx 3 M_\odot$ we estimate that the metallicity inside
a secondary halo corresponding to a blow-away event driven by $N$ supernovae is
\begin{equation}
\label{z}
Z \simeq 1.7\times 10^{-4} N^{2/5}\left({1+z\over 30}\right)^{18/5} (\Omega_b h^2)^{-2/5},
\end{equation}
or $Z \simeq 0.05 N^{2/5} Z_\odot$ at $z\approx 30$ and $(\Omega_b h^2)=
0.05$. This estimate shows that Pop III objects might be responsible for the
origin of polluted regions of relatively high metallicity, althougth of rather small (sub-kpc)
size. Hence, the enrichment of the universe might very well have been very patchy
in its early phases.  

\section{Cosmological Lensing  of  Pop~III SNe}

Pop III supernova events as the ones discussed in the previous Section should
represent one of  the best tracers of star formation at (very) high redshift. The typical
luminosity of a Type II SN is of  order $10^{42}$~erg~s$^{-1}$; the luminosity
of a $10^6 M_\odot$ Pop III object, assuming a reasonable mass-to-light ratio, is
about 100-1000 times smaller. In addition to what we can learn on the formation
of the first objects, Pop III SNe can provide crucial information on cosmological
models due to their different predictions concerning the gravitational lensing
magnification patterns by the intervening matter in the universe.

To investigate this and other aspects, we have have recently undertaken a 
project aimed at simulating the magnification properties of SNe located at
redshift $z_s$, and lensed by the matter distribution predicted by three 
different cosmological models: SCDM (Standard Cold Dark Matter: 
$\Omega_M=1$), LCDM (Lambda Cold Dark Matter: $\Omega_M=0.4,
\Omega_\Lambda=0.6$), CHDM (Cold Hot Dark Matter: $\Omega_M=1
\Omega_\nu=0.3$); all models have $h=0.65$ and are COBE normalized. 
The detailed description of the models, 
simulations and results are presented in Marri \& Ferrara (1998); here
we concentrate on some aspects concerning the magnification probabilities 
of SNe. Technically speaking, the matter distribution in the universe behaves
like a thick gravitational lens. Such 3D distribution must be approximated as 
a sequence of planes, on which the matter is projected, separated by a given
redshift interval (typically $\Delta z = 0.25$). We assume that the lensing matter is
all contained in collapsed point-like dark halos (we neglect baryons at this stage)
whose number density as a function of mass and redshift is calculated
using the Press-Schechter formalism; obviously, this quantity depends on the 
given cosmological model. Such lenses are then distributed randomly on each
plane, \ie neglecting any possible clustering effect.

Our numerical outputs are a set of 
magnification maps displaying the magnification, $\mu$  (\ie the source flux 
enhancement factor) of a point source
located at a given spatial position inside the considered $5' \times 5'$
field of view. There are several unambiguous differences among the
three families of cosmological models that we can identify by analysing
the magnification maps. SCDM models give intense, but not very 
numerous  caustics ($\mu \approx 30-40$) for $z_s \approx 3$; intermediate
magnification caustics appear at higher $z_s$ due to lower mass objects that
have not yet hierarchically merged into larger ones. LCDM models produce
the most intense caustics ($\mu \approx 50$), but most of the map is covered
with more diffuse and lower $\mu$ magnification patterns. Finally, CHDM models,
which form large structures later than $z=3$, show very rare intense events
and many moderate $\mu$ caustics.
We have performed extensive statistical analysis of the magnification maps.
Here we present some results for the cumulative source magnification probability, 
$P(>\mu)$, that is the probability that a source in our field is magnified by more
than a value $\mu$. In terms of this distribution, we can define two useful 
quantities: (i) $\mu_{10}$, or the largest magnification with a probability
larger than 10\%, and (ii) $P(\mu > 10)$, or the probability of magnification
larger than 10 (high magnification probability). Fig. 1 shows the evolution 
of $\mu_{10}$ (top panel) and  $P(\mu > 10)$ (bottom), with  the source redshift.
First, we note that magnifications higher than $\mu = 2$ are very likely
($> 10$\%) for all the cosmological models considered; this fact strongly
enhances the chance of detection of high redshift SNe (Marri \& Ferrara 1998). 
In addition, magnification in SCDM and LCDM models shows appreciable evolution even at 
redshift larger than 5, whereas CHDM magnifications saturate at that epoch, since little 
action is going on at higher redshifts in that particular model. Hence, gravitational
lensing of very high $z$ sources can prove to be a very sharp tool to discriminate
among different cosmological models. By inspecting the bottom panel of Fig. 1,
we also see that there is more than 1\% chance to obtain magnifications $> 10$
at high redshift for the three models. A caveat to keep in mind is that this value 
might be somewhat overestimated by our assumption of point-like lenses and
underestimated if clustering is present. Future work will include a more realistic
schematization of these variables.

\begin{figure}
\vspace{8cm}
\caption{Evolution of the largest magnification with probability larger than 10\%,
$\mu_{10}$, and of the probability of magnification larger than 10, $P(\mu > 10)$,
as a function of the source redshift for the three different cosmological models
considered.}
\end{figure}

\section{Important Points}

We summarize here the main points of the paper:

\begin{itemize}
\item Pop III objects form at $z \approx 30$, provide the first light after CMB and
start to reionize the universe. They also {\it partially} photodissociate the relic $H_2$,
but the negative feedback on galaxy formation, responsible for a
temporary halt in the galaxy formation sequence, is probably
not taking place, as discussed by Ciardi \etal (1998).

\item Blow-away of Pop III seems unavoidable if their masses are 
of the order of $10^6 M_\odot$. Metal enriched gas is cast into the
IGM, producing regions of relatively high ($Z > 10^{-2} Z_\odot$)
metallicity which, however are distributed with a patchy pattern.

\item High redshift SNe in PopIII  lensed by the intervening cosmological 
matter distribution have considerable 
chance ( $> 1$\%) to get magnified by more than 10 times. Thus, detection
chances of these primordial objects are enhanced; at the same time,
lensing of high-$z$ objects will likely allow to discriminate among 
cosmological models.
\end{itemize}


\begin{references}

\reference Ciardi, B.,  \& Ferrara, A. 1997, \apj, 483, L5 

\reference Ciardi, B.,  Abel, T. \& Ferrara, A. 1998, in preparation

\reference Marri, S.,  \& Ferrara, A. 1998, in preparation

\reference Haiman, Z., Rees, M. J., \& Loeb, A. 1997, \apj, 484, 985

\reference Tegmark, M., Silk, J., Rees, M.J., Blanchard, A., Abel, T. \& Palla, F.
1997, \apj, 474, 1
\end{references}
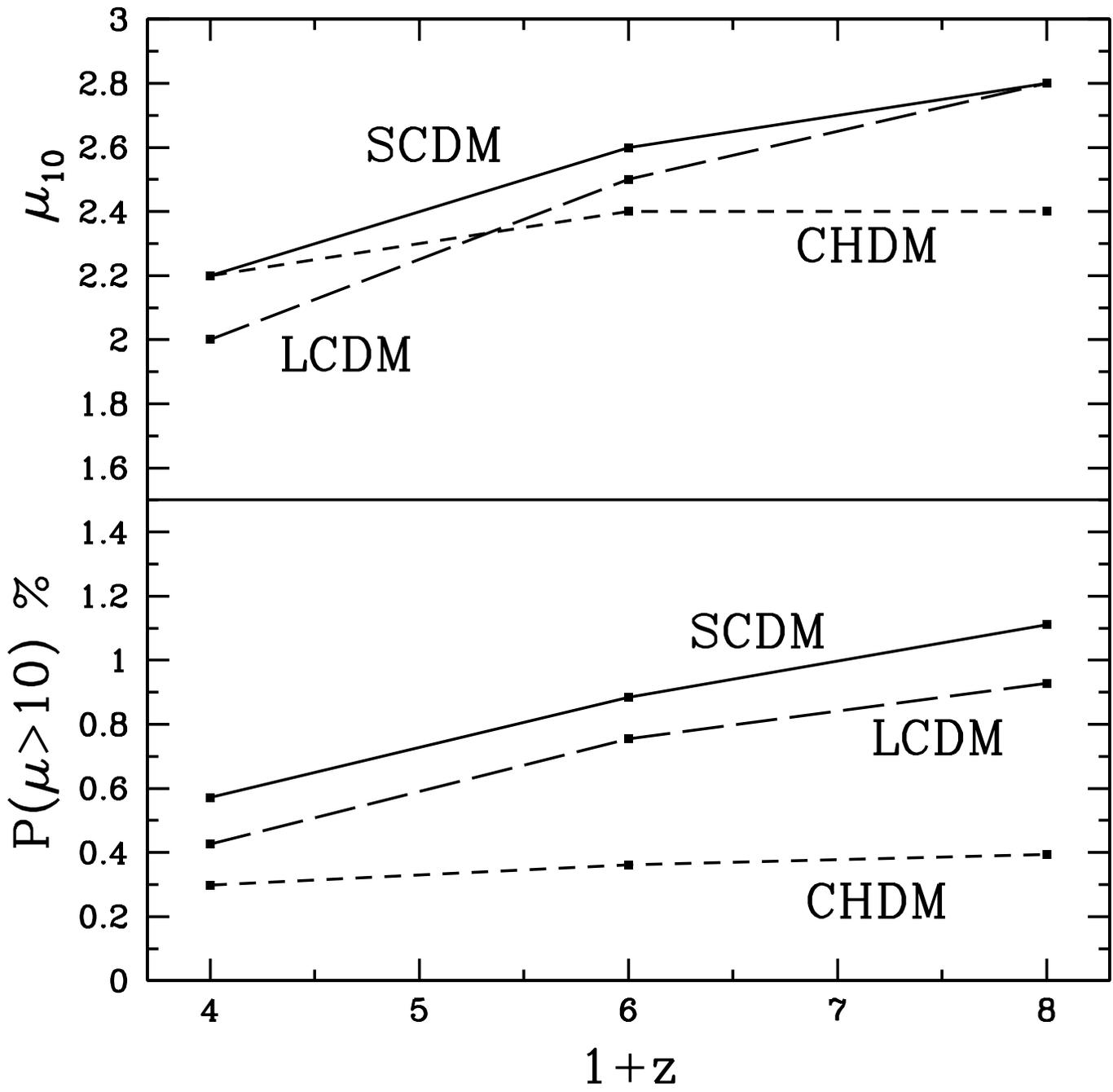
\end{document}